\documentstyle[times,epsfig]{jaa}

\begin{document}

\title[Angular momentum transport in accretion disks]
{Angular momentum transport in quasi-Keplerian accretion disks} 

\author[Subramanian, Pujari, \& Becker]{Prasad Subramanian,$^1$
   \thanks{e-mail: psubrama@iucaa.ernet.in} B. S. Pujari, $^2$
   \thanks{e-mail: bspujari@rediffmail.com}, Peter A. Becker, $^3$
   \thanks{e-mail: pbecker@gmu.edu}\\
 $^1$Inter-University Center for Astronomy and Astrophysics, P.O Bag 4,
Ganeshkhind, Pune - 411 007, India \\ 
 $^2$Dept. of Physics, University of Pune, Pune - 411 007, India \\
$^3$ School of Computational Sciences, George Mason University, Fairfax,
VA 22030, USA}

\maketitle

\begin{abstract}
We reexamine arguments advanced by Hayashi \& Matsuda (2001), who claim
that several simple, physically motivated derivations based on mean free
path theory for calculating the viscous torque in a quasi-Keplerian
accretion disk yield results that are inconsistent with the generally
accepted model. If correct, the ideas proposed by Hayashi \& Matsuda
would radically alter our understanding of the nature of the angular
momentum transport in the disk, which is a central feature of accretion
disk theory. However, in this paper we point out several fallacies in
their arguments and show that there indeed exists a simple derivation
based on mean free path theory that yields an expression for the viscous
torque that is proportional to the radial derivative of the angular
velocity in the accretion disk, as expected. The derivation is based
on the analysis of the epicyclic motion of gas parcels in adjacent eddies
in the disk.
\end{abstract}


\section{Introduction}

Accretion disks around a central compact object are ubiquitous in
astrophysics. In such systems, the plasma accreting onto the central
object typically takes the form of a swirling disk. The disk-like nature
of accretion is due to the angular momentum originally carried by the
accreting material, and also to the spin of the central object, which
provides a preferred axis of rotation. At a given radius in a quasi-Keplerian
accretion disk, the average motion of the gas parcels is an approximately
circular Keplerian orbits, combined with a radial sinking motion as the
parcel gradually moves deeper into the potential well of the central
mass. The radial velocity that is far smaller than the Keplerian
(azimuthal) velocity. A detailed description of quasi-Keplerian accretion
disks can be found in Pringle (1981), among several other references.

The specific angular momentum (angular momentum per unit mass) carried
by a parcel of gas in a quasi-Keplerian accretion disk is very close to
the Keplerian value $\sqrt{G M R}$, where $R$ is the radius measured
from the central object with mass $M$. It follows that a parcel of
plasma has to lose angular momentum in order to sink from a larger orbit
into a smaller one and eventually cross the event horizon. In the
absence of strong winds or jets, the angular momentum must therefore
flow {\it outwards} in such an accretion disk in order to enable
accretion to proceed. In disks of turbulent fluid, the flow of angular
momentum is due to the exchange of plasma parcels in neighboring annuli
with different angular velocities. This fluid picture may describe
certain astrophysical cases such as the rings of Saturn, but in many
other cases of astrophysical interest, the coupling between adjacent
annuli is provided not by ``parcel interchange,'' but rather by the
magnetic field. In either the fluid or magnetohydrodynamical (MHD)
scenarios, the angular momentum flow is expressed as a torque between
neighboring annuli. Our focus here is on the fluid picture, and on the
validity of the various attempts to analyze quantitatively the angular
momentum transport associated with the interchange of parcels.

There are several previous papers in which derivations of the
viscous torque in fluid dynamical situations are developed based on
simple, physically motivated arguments (e.g., Frank, King, \& Raine 1985,
1992, 2002; Hartmann 1998). However, some of these do not yield an
expression for the viscous torque that is proportional to the radial
derivative of the angular velocity. This is a problem because the
viscosity is fundamentally due to the ``rubbing'' of matter in adjacent
radial annuli in the disk, and consequently the viscosity should {\it
vanish} in the case of solid body rotation with $\Omega(R)=\rm
constant$, where $\Omega(R)$ is the angular velocity in the disk at
radius $R$. Hayashi \& Matsuda (2001; hereafter HM) recognized this
point, and attempted to clear up some of the confusion by carefully
examining the previously published derivations of the viscosity that
were based on the mean free path approach. They concluded that the mean
free path approach inevitably leads to an inward rather than outward
flow of angular momentum in the disk, which is unphysical. However, we
argue that the reasoning of HM was flawed because they did not consider
the epicyclic nature of the parcel trajectories. We present a simple
derivation based on mean free path theory, combined with the actual
epicyclic motion of the gas parcels, that in fact yields a physically
reasonable expression for the viscous torque between neighboring annuli
in a quasi-Keplerian accretion disk.

We discuss the standard fluid dynamical formulation of viscous torques in
an accretion disk in \S~2. We then discuss the problems with the various
previous derivations that attempted to utilize a mean free path approach
to compute the viscous torque in \S~3. In \S~4 we present a simple physical
derivation based on analysis of the epicyclic (ballistic) motion of
gas parcels in adjacent eddies, and we demonstrate that this approach yields
the expected form for the torque in terms of the gradient of the angular
velocity. We present our final conclusions in \S~5.

\section{Standard fluid dynamics treatment of viscous torque}

The equation of motion for a viscous, incompressible fluid can be
written as (Landau \& Lifshitz 1987)
\begin{equation}
\frac{\partial \, \vec{v}}{\partial\,t} + (\vec{v}\cdot\vec{\nabla})\,\vec{v}
= \frac{\vec{F}}{\rho} \, ,
\label{eq1}
\end{equation}
where the force density $\vec{F}$ is defined via the viscous stress tensor
$\sigma_{ik}$ as follows:
\begin{eqnarray}
\nonumber
F_{i} = \frac{\partial\,\sigma_{ik}}{\partial\,x_{k}} \, ,\\
\sigma_{ik} = - p\,\delta_{ik} + \eta\,\biggl ( \frac{\partial\,v_{i}}
{\partial\,x_{k}} + \frac{\partial\,v_{k}}{\partial\,x_{i}} \biggr ) \, .
\label{eq2}
\end{eqnarray}
The fluid velocity is denoted by $\vec{v}$ and its density and pressure
by $\rho$ and $p$, respectively. The coefficient of dynamic viscosity (g
cm$^{-1}$ s$^{-1}$ in cgs units) is denoted by $\eta$ and we have used
the Einstein summation convention in equation~(\ref{eq2}). For the special case
of a thin, azimuthally symmetric accretion disk, the only non-negligible
component of the viscous stress tensor in cylindrical coordinates ($R$,
$\phi$, $z$) is
\begin{equation}
\sigma_{R\,\phi} = - \eta\,R\,\frac{\partial\,\Omega}{\partial R}\, .
\label{eq3}
\end{equation}
The viscous stress (force per unit area) is thus directly proportional
to the radial derivative of the angular velocity. This is by far the
cleanest and most rigorous way to derive the azimuthal equation of
motion for a quasi-Keplerian accretion disk (see, for example, chapter 1
of Subramanian 1997). It is implicity assumed that the dynamic viscosity
$\eta$ arises out of local effects; i.e., due to momentum exchange
between neighboring annuli of the accretion disk, as with molecular
viscosity. It is well known (Shakura \& Sunyaev 1973; Pringle 1981) that
molecular viscosity is far too small to account for angular momentum
transport in accretion disks around active galactic nuclei. Identifying
suitable candidates for the microphysical viscosity mechanism operative
in such disks is a subject of intensive research.

\section{Simplified treatments of viscous torque in accretion disks}

Pringle (1981) derived the azimuthal component of the equation of motion
in accretion disks starting from first principles in a simple,
physically motivated manner. The equation reads as follows:
\begin{equation}
\frac{\partial\,(\Sigma\,R^{3}\,\Omega)}{\partial\,t}
+ \frac{\partial\,(\Sigma\,v_{_{R}} \, R^{3}\,\Omega)}{\partial\,R}
= - \frac{1}{2 \pi}\frac{\partial {\cal G}}{\partial R} \, ,
\label{eq4}
\end{equation}
where $\Sigma$ is the surface density of plasma in the disk, $v_{_R} < 0$
is the radial accretion velocity, and ${\cal G} > 0$ is the torque exerted by
the material inside radius $R$ on the material outside that radius.
Throughout the remainder of the paper, we shall assume that the disk
has a steady-state (time-independent) stucture, although this is not
essential for our results. In the standard approach introduced by Shakura
\& Sunyaev (1973) and adopted by Pringle (1981), the torque ${\cal G}$ is related
to $\Omega$ via
\begin{equation}
{\cal G} = 4 \pi R^2 H \sigma_{R\,\phi}
= -2 \pi R^3 \Sigma \, \nu {d\Omega\over d R}
\label{eq5}
\, ,
\end{equation}
where $H(R)$ is the half-thickness of the disk at radius $R$ and
$\nu=\eta/\rho$ is the kinematic viscosity coefficient. The stress and
torque are therefore proportional to $d\Omega/d R$, in agreement with
equation~(\ref{eq3}). This prescription has been applied in many disk
stucture calculations. In particular, it has been shown recently (Becker
\& Le 2003) that fully relativistic and self-consistent models for hot,
advection-dominated accretion disks can be constructed by applying the
Shakura-Sunyaev viscosity prescription throughout the entire disk,
including the region close to the event horizon.

A number of authors have attempted to confirm the general form of
equation~(\ref{eq5}) by using simple physical arguments. However,
several of these derivations have errors in them, and they do not always
result in an expression for ${\cal G}$ that is proportional to $d\Omega/dR$, as
pointed out by HM. We briefly review the relevant derivations below, and
we also point out errors in the approach adopted by HM. We then present
a new, heuristic derivation of equation~(\ref{eq5}) that is based on a
careful analysis of the ballistic motion of two parcels as they exchange
radii. This derivation leads to the expected conclusion that ${\cal G}
\propto d\Omega/dR$.

Although the argument given by HM is rather indirect, their main point
can be understood through a simple examination of the angular momentum
transport resulting from the interchange of fluid elements in a disk
that is rotating as a solid body, i.e., with $\Omega(R)=\Omega_0=\rm
constant$. In this case, the angular momentum per unit mass, denoted by
$J\equiv R^2\,\Omega(R)$, is given by $J = R^2 \, \Omega_0$, and this
quantity increases rather strongly as a function of the radius $R$.
Hence if two parcels of fluid on opposite sides of radius $R$ were
exchanged due to some turbulent or convective process, then clearly
angular momentum would be transported in the {\it inward} direction,
since the blob that was originally outside the annulus will have more
angular momentum than the interior blob. However, this result is
unphysical, because in the case of solid body rotation, there is no
``rubbing'' between adjacent fluid annuli, and therefore there should be
no torque and no angular momentum transport. Any successful microphysical
model for the angular momentum transport in the disk based on mean free path
theory must somehow resolve this apparent paradox. In the following
sections, we provide a detailed consideration of the reasoning employed
in the previous published derivations, including that of HM, and we
conclude that when properly carried out, the mean free path approach can
yield a result for the viscous torque that correctly vanishes in the
case of solid body rotation.

\subsection{Derivation of ${\cal G}$ by Hartmann (1998)}

The specific angular momentum $J_{\rm in}$ carried by material originating
at a radius $R - \lambda/2$ in a quasi-Keplerian accretion disk is given
by Hartmann (1998) as $J_{\rm in} = (R - \lambda/2)\,\Omega(R - \lambda/2)
\, ,$ where $\lambda$ is the mean free path over which parcels of plasma
exchange angular momentum. As pointed out by HM, this expression is
incorrect, and the correct expression should read as follows:
\begin{equation}
J_{\rm in} = \left(R - \frac{\lambda}{2}\right)^{2}\,
\Omega\left(R - \frac{\lambda}{2}\right) \, .
\label{eq6}
\end{equation}
One can write an analogous expression for $J_{\rm out}$, the specific
angular momentum carried by material originating a a radius $R +
\lambda/2$, by reversing the sign of $\lambda$. As shown by HM, if one
expands $\Omega(R - \lambda/2)$ to first order in $\lambda$ as $\Omega(R
- \lambda/2) \sim \Omega(R) - (\lambda/2)(d\Omega/dR)$, the result obtained
for the difference between the specific angular momenta is
\begin{equation}
J_{\rm in} - J_{\rm out} = -\, \lambda\,\frac{d}{dR}\,(R^{2}\,\Omega) \, .
\label{eq7}
\end{equation}
The characteristic time for the interchange of the matter between the
two radii is $\Delta t = \lambda/w$, where $w$ is the turbulent velocity
of the fluid parcels. The total mass of fluid involved in the interchange
is $\Delta M = 2 \pi R \lambda \Sigma$, and it follows that the net rate of
flow of angular momentum from the inner ring at $R - \lambda/2$ towards the
outer ring at $R + \lambda/2$ (or equivalently, the viscous torque exerted
by the ring at $R - \lambda/2$ on the ring at $R + \lambda/2$) is
\begin{equation}
{\cal G} = {\Delta M \over \Delta t}\,
(J_{\rm in} - J_{\rm out})
= 2 \pi R\,\Sigma\,w\,(J_{\rm in} - J_{\rm out}) \ ,
\label{eq8}
\end{equation}
or
\begin{equation}
{\cal G} = -\, 2 \pi R\,\Sigma\,
\beta \, \nu\,\frac{d}{dR}\,(R^{2}\,\Omega) \, ,
\label{eq8b}
\end{equation}
where we have set $\nu = w \lambda / \beta$, with $\beta$ denoting a
constant of order unity. The viscous torque in a quasi-Keplerian
accretion disk should tend to smooth out gradients in the angular
velocity so as to attain solid body rotation ($d\Omega/dR = 0$).
However, the expression for the viscous torque in equation~(\ref{eq8b})
is such that it tends to attain a flow with {\it constant angular
momentum}, i.e., the ``equilibrium'' condition is $d(R^{2}\,\Omega) /dR
= 0$. As HM point out, this expression is therefore unphysical.

\subsection{Derivation of ${\cal G}$ by Frank, King, \& Raine (1992)}

We next turn our attention to another derivation of the viscous torque
given by Frank, King, \& Raine (1992). At the heart of their derivation
is the claim that the linear velocity of material at radius $R -
\lambda/2$ as seen by an observer situated at radius $R$ is given by
\begin{equation}
v_{\rm rel}\left(R - \frac{\lambda}{2}\right)
= \left(R-\frac{\lambda}{2}\right)\,
\Omega\left(R-\frac{\lambda}{2}\right)
+ \Omega(R)\,\frac{\lambda}{2} \, .
\label{eq9}
\end{equation}
Based on this, one can write an expression for $L_{\rm in}$, the rate
at which angular momentum crosses an annulus at radius $R$ in the direction
of increasing $R$, as
\begin{eqnarray}
\nonumber
L_{\rm in} = 2 \pi R \, \Sigma \, w \, \left(R - \frac{\lambda}{2}\right)
\, v_{\rm rel} \left(R - \frac{\lambda}{2}\right) \\
\nonumber
= 2 \pi R \, \Sigma \, w \, \left(R - \frac{\lambda}{2}\right)
\, \left[ \left(R - \frac{\lambda}{2}\right)\,
\Omega\left(R - \frac{\lambda}{2}\right)
+ \Omega(R)\,\frac{\lambda}{2} \right] \\
\simeq  2 \pi R \, \Sigma \, w \, \left(R - \frac{\lambda}{2}\right)
\, \left[ R\,\Omega(R) - R\,\frac{\lambda}{2}\,\frac{d\,\Omega}{d\,R}
\right] \, ,
\label{eq10}
\end{eqnarray}
where we have expanded $\Omega(R - \lambda/2)$ to first order in
$\lambda$ to arrive at the final expression. We can write an analogous
expression for $L_{\rm out}$, the rate at which angular momentum crosses
radius $R$ in the inward direction, by reversing the sign of $\lambda$.
To first order in $\lambda$, the torque exerted on the plasma outside
radius $R$ by the plasma inside that radius is then
\begin{equation}
{\cal G} = L_{\rm in} - L_{\rm out} = - 2 \pi R^2 \, \beta
\, \nu \, \Sigma \, \frac{d}{dR}\,(R\,\Omega) \, .
\label{eq11}
\end{equation}
Since the right-hand side is proportional to $d(R\,\Omega)/dR$,
in this case the torque will lead to a uniform {\it linear velocity},
and not to a uniform {\it angular velocity} as we require on physical
grounds. As HM point out, this expression for ${\cal G}$ is therefore also
incorrect.

\subsection{Correction proposed by HM}

HM claim that this is because Frank, King, \& Raine (1992) have used
an incorrect expression for $v_{\rm rel}$ (i.e., eq.~[\ref{eq9}]). They
assert that the linear velocity of the plasma at $(R - \lambda/2)$ as
viewed by an observer at radius $R$ should instead be given by
\begin{equation}
v_{\rm rel}\left(R - \frac{\lambda}{2}\right)
= \left(R-\frac{\lambda}{2}\right)\,
\Omega\left(R-\frac{\lambda}{2}\right)
- R\,\Omega(R) + \Omega(R)\,\frac{\lambda}{2} \, .
\label{eq12}
\end{equation}
By employing this expression for $v_{\rm rel}$ and following the same
procedure used to obtain equation~(\ref{eq11}), they find that to first
order in $\lambda$,
\begin{equation}
{\cal G} = L_{\rm in} - L_{\rm out} = - 2 \pi R^3 \, \beta
\, \nu \, \Sigma \, \frac{d\Omega}{dR} \, .
\label{eq13}
\end{equation}
This expression for ${\cal G}$ does indeed have the correct dependence on
$d\Omega/dR$, but nonetheless we claim that equation~(\ref{eq12})
for the relative velocity used by HM is incorrect. The correct
expressions for the relative velocities are in fact (see, e.g.,
Mihalas \& Binney 1981)
\begin{eqnarray}
\nonumber
v_{\rm rel}\left(R - \frac{\lambda}{2}\right)
= \left(R-\frac{\lambda}{2}\right)\,
\Omega\left(R-\frac{\lambda}{2}\right) - R\,\Omega(R)\, \\
v_{\rm rel}\left(R + \frac{\lambda}{2}\right)
= \left(R+\frac{\lambda}{2}\right)\,
\Omega\left(R+\frac{\lambda}{2}\right) - R\,\Omega(R) \, ,
\label{eq14}
\end{eqnarray}
where $v_{\rm rel}(R - \lambda/2)$ denotes the velocity of a plasma parcel
at $R-\lambda/2$ as seen by an observer at $R$, and and $v_{\rm rel}(R +
\lambda/2)$ denotes the velocity of a plasma parcel at $R + \lambda/2$
as seen by an observer at $R$. In a quasi-Keplerian accretion disk,
$\Omega \propto R^{-3/2}$, and therefore the first velocity is positive
and the second is negative. These expressions for the relative
velocities assume that the plasma parcels at $R$, $R - \lambda/2$, and $R
+ \lambda/2$ all lie on the same radial line through the central object. If
this is not true, then the expressions will contain additional terms, as
Mihalas \& Binney (1981) show in the context of the relative velocity
between the sun and stars in our galaxy. Since we are dealing with
material transport over lengths of the order of $\lambda$ that are very
small in comparison with $R$, this assumption is quite valid in our
accretion disk application. Using equation~(\ref{eq14}) for the relative
velocities, we now obtain
\begin{eqnarray}
\nonumber
L_{\rm in} = 2 \pi R \, \Sigma \, w\,\left(R - \frac{\lambda}{2}\right)
\, \left[ \left(R - \frac{\lambda}{2}\right)
\,\Omega\left(R - \frac{\lambda}{2}\right) - R\,\Omega(R) \right] \\
\nonumber
\simeq - 2 \pi R^2 \, \Sigma\,w\,\frac{\lambda}{2} \,\left[R\,\frac{d\Omega}
{dR} + \Omega(R) \right] \\
\nonumber
L_{\rm out} = 2 \pi R \, \Sigma \, w\,\left(R + \frac{\lambda}{2}\right)
\, \left[\left(R + \frac{\lambda}{2}\right)
\,\Omega\left(R + \frac{\lambda}{2}\right) - R\,\Omega(R) \right]\\
\simeq \, 2 \pi R^2 \, \Sigma\,w\,\frac{\lambda}{2} \,
\left[R\,\frac{d\Omega}{dR} + \Omega(R) \right] \, ,
\label{eq15}
\end{eqnarray}
where the final expressions for $L_{\rm in}$ and $L_{\rm out}$ are
correct to first order in $\lambda$. We thus obtain for the viscous
torque
\begin{equation}
{\cal G} = L_{\rm in} - L_{\rm out}
= - \, 2 \pi R^2 \, \Sigma \ \beta \, \nu {d \over dR}
\, (R \, \Omega) \ .
\label{eq15.5}
\end{equation}
This result is identical to equation~(\ref{eq11}), and therefore
it too is incorrect.

\section{Derivation based on epicyclic parcel motion}

\begin{figure}
\begin{center}
\epsfig{file=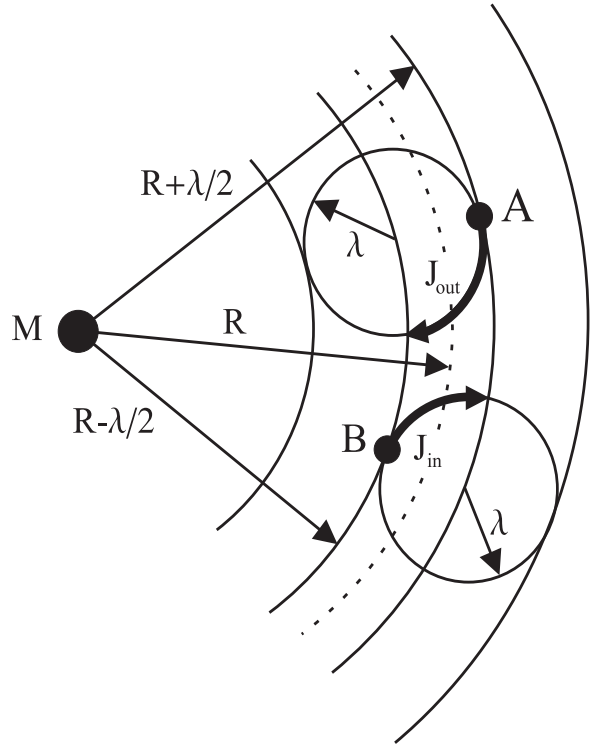,height=10cm,width=7.8cm}
\caption{Angular momentum transport in the disk involves the interchange
of two parcels, A and B, initally located at radii $R + \lambda/2$ and
$R - \lambda/2$, respectively, around central mass $M$. They each
participate in ballistic, epicyclic motion in eddies of radius $\lambda$,
with A moving inward and B moving outward. See the discussion in the
text.}
\end{center}
\end{figure}

We have demonstrated in \S~3 that due to various errors, the previous
derivations of the viscous torque based on mean free path theory do not
yield results consistent with the classical theory of viscous transport.
We shall now focus on a heuristic analysis of the angular momentum transport
that combines the mean free path approach with a proper treatment of the
epicyclic motion of parcels of gas in adjacent eddies in the accretion disk.
The physical picture is presented in Fig.~1. Let us suppose that $\lambda$
represents the mean distance a parcel travels freely before having its motion
disrupted by interaction with the surrounding material in the disk. The
disruption in this case involves hydrodynamical interaction, and therefore
$\lambda$ is the damping length for the turbulence as well as the mean free
path for the parcel motion. It follows that the turbulence in the disk can
be characterized by whirls with a mean radius $\lambda$.

Next we focus on the interchange of two gas parcels of unit mass, A and
B, which are initially located at radii $R + \lambda/2$ and $R -
\lambda/2$, respectively (see Fig.~1). The two parcels each ``ride''
turbulent eddies with radius $\lambda$, and they start out at the outer
(parcel A) or inner (parcel B) edge of their respective eddies. Hence
each begins its motion with zero radial velocity, at a turning point in
its orbit. Parcel A moves in the inward direction, and therefore its
angular momentum is sub-Keplerian compared with the average disk at
radius $R + \lambda/2$. Similarly, parcel B is super-Keplerian compared
with the mean disk at radius $R - \lambda/2$ and therefore it moves in
the outward direction. The motion of the parcels is epicyclic as viewed
from the reference frame of an observer who remains at the starting
radius and travels along with the Keplerian angular velocity. Since
$\lambda$ is the mean damping length for the turbulence, each of the
parcels will experience ballistic motion over a radial length scale
comparable to $\lambda$, after which the ballistic motion (and the
parcel itself) will be absorbed by the surrounding gas. It
follows that, on average, parcel A will deposit its angular momentum
($J_{\rm out}$) at radius $R - \lambda/2$, and parcel B will deposit its
angular momentum ($J_{\rm in}$) at radius $R + \lambda/2$. Hence the two
parcels exchange radii during the process. We shall proceed to compute
the net angular momentum exchange, $J_{\rm in} - J_{\rm out}$, by
considering the motion of these parcels in detail.

Gas parcel A participates in the motion of turbulent eddy A, which has
its center located at radius $R - \lambda/2$ and possesses turning
points at radii $R + \lambda/2$ and $R - 3 \lambda/2$. Likewise, parcel
B experiences the motion of eddy B, with its center located at radius $R
+ \lambda/2$, and with turning points a radii $R + 3 \lambda/2$ and $R -
\lambda/2$. The specific energy $E$ and the specific angular momentum
$J$ of each parcel is conserved during the ballistic phase of its motion,
until it travels a mean distance $\lambda$ and is damped by the surrounding
material. For a particle moving ballistically in the Newtonian gravitational
field of a central mass $M$, the quantities $E$ and $J$ are related by the
classical energy equation
\begin{equation}
E = {1 \over 2} \, v_{_R}^2 + {1 \over 2} \, {J^2 \over R^2}
- {GM \over R} \ ,
\label{energy1}
\end{equation}
where $v_{_R}$ is the radial component of the velocity. Turning
points in the radial motion occur where $v_{_R}$ vanishes, so that
we can write
\begin{equation}
E = {1 \over 2} \, {J^2 \over R_1^2} - {GM \over R_1}
= {1 \over 2} \, {J^2 \over R_2^2} - {GM \over R_2}\ ,
\label{energy2}
\end{equation}
where $R_1$ and $R_2$ denote the two turning point radii. This equation
can be easily solved for the angular momentum $J$ as a function of $R_1$
and $R_2$. The result obtained is
\begin{equation}
J = \left({2 \, G M \, R_1 \, R_2 \over R_1 + R_2}\right)^{1/2} \ .
\label{eq16}
\end{equation}
We can use equation~(\ref{eq16}) to conclude that the
angular momentum of parcel A is equal to
\begin{equation}
J_{\rm out} = \sqrt{G M R} \, \left(1 + {1 \over 2}{\lambda \over R}\right)
\left(1 - {3 \over 2}{\lambda \over R}\right)^{1/2}
\left(1 - {1 \over 4}{\lambda^2 \over R^2}\right)^{-1/2} \ .
\label{eq17}
\end{equation}
Likewise, the angular momentum of parcel B is given by
\begin{equation}
J_{\rm in} = \sqrt{G M R} \, \left(1 - {1 \over 2}{\lambda \over R}\right)
\left(1 + {3 \over 2}{\lambda \over R}\right)^{1/2}
\left(1 - {1 \over 4}{\lambda^2 \over R^2}\right)^{-1/2} \ .
\label{eq18}
\end{equation}

In the spirit of the mean free path approach, we are interested
in computing the value of the net angular momentum transport,
$J_{\rm in} - J_{\rm out}$, to first order in the small parameter
$\lambda/R$. The corresponding results obtained for $J_{\rm in}$
and $J_{\rm out}$ are
\begin{equation}
J_{\rm out} = \sqrt{G M R} \, \left(1 - {1 \over 4}{\lambda \over R}
\right) + {\rm O}\left[{\lambda^2 \over R^2}\right]
\ ,
\label{eq19}
\end{equation}
\begin{equation}
J_{\rm in} = \sqrt{G M R} \, \left(1 + {1 \over 4}{\lambda \over R}
\right) + {\rm O}\left[{\lambda^2 \over R^2}\right]
\ .
\label{eq20}
\end{equation}

The angular velocity $\Omega(R)$ in a quasi-Keplerian accretion disk
is very close to the Keplerian value, and therefore we can write
\begin{equation}
\Omega(R) = \sqrt{G M \over R^3}
\ .
\label{eq21}
\end{equation}
It follows that
\begin{equation}
R^2 \, \Omega\left(R + {\lambda \over 6}\right)
= \sqrt{G M R} \, \left(1 + {1 \over 6}{\lambda \over R}
\right)^{-3/2} \ ,
\label{eq22}
\end{equation}
or, to first order in $\lambda/R$,
\begin{equation}
R^2 \, \Omega\left(R + {\lambda \over 6}\right)
= \left(1 - {1 \over 4}{\lambda \over R}
\right) + {\rm O}\left[{\lambda^2 \over R^2}\right]
\ .
\label{eq23}
\end{equation}
This also implies that
\begin{equation}
R^2 \, \Omega\left(R - {\lambda \over 6}\right)
= \left(1 + {1 \over 4}{\lambda \over R}
\right) + {\rm O}\left[{\lambda^2 \over R^2}\right]
\ .
\label{eq24}
\end{equation}
Comparing equations~(\ref{eq19}) and (\ref{eq20}) with equations
(\ref{eq23}) and (\ref{eq24}), we find that to first order in
$\lambda/R$, the net angular momentum transfer is given by
\begin{equation}
J_{\rm in} - J_{\rm out}
= R^2 \, \Omega\left(R - {\lambda \over 6}\right)
- R^2 \, \Omega\left(R + {\lambda \over 6}\right)
\simeq - {\lambda \over 3} \, R^2 \, {d\Omega\over dR} \ .
\label{eq25}
\end{equation}
By following the same steps leading to equation~(\ref{eq8}), we now obtain
\begin{equation}
{\cal G} = 2 \pi R\,\Sigma\,w\,(J_{\rm in} - J_{\rm out})
\simeq - 2 \pi R^3\,\nu\,\Sigma\,\frac{d\Omega}{dR} \, ,
\label{eq26}
\end{equation}
where the final expression is correct to first order in $\lambda$ and
we have set $\nu = w \lambda/3$ so that $\beta=3$. Note that this
result agrees very well with the Shakura-Sunyaev form (eq.~[\ref{eq5}]).
Hence we have demonstated using a simple heuristic derivation that the
viscous torque is indeed proportional to the gradient of the angular
velocity in an accretion disk within the context of a mean free path,
parcel-exchange picture.

\section{Conclusion}

The derivation presented in \S~4 clearly employs mean free path theory,
since we assumed that the fluid parcels travel an average radial
distance $\lambda$ before being damped by the surrounding gas. We have
thus shown that there does exist a simple derivation, based on mean free
path theory, that yields an expression for the viscous torque ${\cal G}$
(eq.~[\ref{eq26}]) that is directly proportional to the radial
derivative of the angular velocity, $d\Omega/dR$, in agreement with our
physical expectation. This expression for ${\cal G}$ can be used in
equation~(\ref{eq4}) to proceed further in deriving the structure of the
accretion disk, as in Pringle (1981). Our results provide a simple but
important unification of the ``parcel interchange'' viscosity model with
the MHD viscosity model, in which the coupling that transports the
angular momentum is provided by the magnetic field rather than by fluid
turbulence (Frank, King, \& Raine 2002). In the MHD model, the torque is
found to be proportional to the gradient of the angular velocity,
$d\Omega/dR$, in agreement with the results we have obtained here by
applying the mean free path approach to the case of fluid turbulence.


\begin{thebibliography}{10}

\bibitem{1} Becker, P. A., Le, T. 2003, {\em Astrophys. J.,} {\bf 588},
408

\bibitem{2} Frank, J., King, A., Raine, D. 1985, {\em Accretion Power
in Astrophysics, 1st edition}, Cambridge University Press

\bibitem{3} Frank, J., King, A., Raine, D. 1992, {\em Accretion Power
in Astrophysics, 2nd edition}, Cambridge University Press

\bibitem{4} Frank, J., King, A., Raine, D. 2002, {\em Accretion Power
in Astrophysics, 3rd edition}, Cambridge University Press

\bibitem{5} Hartmann, L. 1998, {\em Accretion Processes in Star Formation},
Springer

\bibitem{6} Hayashi, E., Matsuda, T. 2001, {\em Prog. Theor. Phys.,}
{\bf 105}, 531 (HM)

\bibitem{7} Landau, L. D., Lifshitz, E. M. 1987, {\em Fluid Mechanics,
2nd edition}, Pergamon Press

\bibitem{8} Mihalas, D., Binney, J. 1981, {\em Galactic Astronomy, 2nd edition},
W. H. Freeman \& Co., p. 468

\bibitem{9} Pringle, J. 1981, {\em Ann. Rev. Astron. Astrophys.,} {\bf 19},
137

\bibitem{10} Shakura, N. I., Sunyaev, R. I. 1973, {\em Astron. Astrophys.,}
{\bf 24}, 337

\bibitem{11} Subramanian, P. 1997, {\em Ph.D. Thesis, George Mason University}

\end{thebibliography}
\end{document}